\documentclass{article}

\usepackage[final]{neurips_2020}

\usepackage[utf8]{inputenc} 
\usepackage[T1]{fontenc}    
\usepackage{hyperref}       
\usepackage{url}            
\usepackage{booktabs}       
\usepackage{amsfonts}       
\usepackage{nicefrac}       
\usepackage{microtype}      
\usepackage{graphicx}
\usepackage{caption, subcaption}

\title{Explainable Link Prediction \\ for Privacy-Preserving Contact Tracing}

\author{
  Balaji Ganesan \\
  IBM Research India \\
  Bengaluru, India \\
  \texttt{bganesa1@in.ibm.com} \\
  \And
  Hima Patel \\
  IBM Research India \\
  Bengaluru, India \\
  \texttt{himapatel@in.ibm.com} \\
  \And
  Sameep Mehta \\
  IBM Research India \\
  Bengaluru, India \\
  \texttt{sameepmehta@in.ibm.com} \\
}

\begin{document}

\maketitle

\begin{abstract}
Contact Tracing has been used to identify people who were in close proximity to those infected with SARS-Cov2 coronavirus. A number of digital contract tracing applications have been introduced to facilitate or complement physical contact tracing. However, there are a number of privacy issues in the implementation of contract tracing applications, which make people reluctant to install or update their infection status on these applications. In this concept paper, we present ideas from Graph Neural Networks and explainability, that could improve trust in these applications, and encourage adoption by people.
\end{abstract}

\section{Introduction}

Contact Tracing is the task of searching people who might have come in physical contact with individuals of interest, like those who have tested positive for SARS-Cov2 coronavirus. Digital Contact Tracing is the version of contact tracing which relies on mobile phones and/or wearable devices for monitoring events when individuals were in close proximity.

As governments have tried to encourage the use of such digital contact tracing applications in response to COVID19, a number of privacy related issues were raised. \cite{bengio2020need} proposed a number of recommendations for contact tracing applications. Although a number of solutions have been proposed including those based on federated learning, the adoption of contact tracing apps by people has been lukewarm. As shown by \cite{mit2020covidtracingtracker2}, the adoption is less than 15\% of the population in several countries that have seen significant COVID19 outbreaks.

The low adoption of contact tracing and the related exposure notification apps have lead to concerns that these apps are not going to work if people are not motivated to use them \cite{bengio2020b}. Even among the people who have downloaded such apps, the usage remains particularly low, making them ineffective in combating the spread of SARS-Cov2 virus. \cite{abueg2020modeling} argued that contact tracing apps could be useful even when the adoption is low. But increased adoption and information sharing benefits the society at large.

\begin{figure}
  \includegraphics[width=\columnwidth]{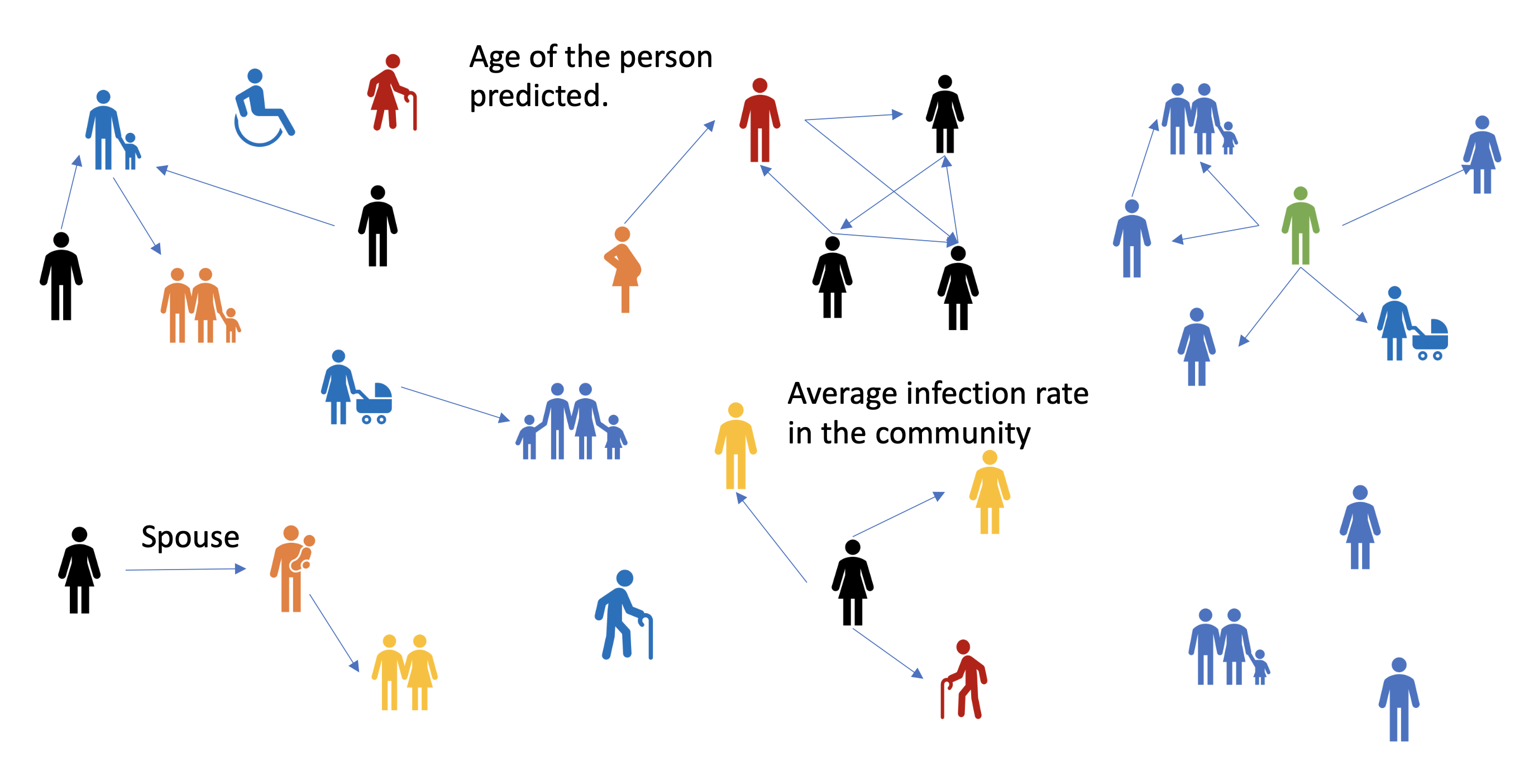}
  \caption{Contact tracing of COVID19 positive persons}
  \label{fig:watchlist}
\end{figure}

In this work, we propose ideas that can encourage the adoption of digital contact tracing and exposure notification applications, while adhering to the privacy considerations. First, we propose that not all instances of proximity between individuals need to generate exposure notifications. Instead only predictions of possible exposure to the coronavirus should lead to alerts. This can be accomplished using link prediction and node classification tasks in graphs.

Link Prediction is the task of finding missing links in a graph. Given a property graph where nodes are people, and their physical contacts are links, Graph Neural Network (GNN) models can be trained to predict additional \textit{exposure links}. These links can happen even when there is no recorded physical proximity event (because of apps being switched off or people not carrying the phone on them during a chat in office and other similar potential exposure events). If we are able to predict exposure links well, classifying whether a node is exposed, is a relatively easier problem. So in this work, we focus only on link prediction. The other related task in graphs, namely entity resolution has applications in physical contact tracing and the solutions proposed here are applicable for entity resolution as well.

Explaining the neural model predictions is naturally very important to build trust in the digital contact tracing apps. But making the explanations more human understandable is particularly important in applications aimed at the general population. We propose an improvement to the state of the art Anchors solution of \cite{ribeiro2018anchors}, and also introduce a new path ranking based explainability solution. Like in any social network, the contact tracing apps are only as good as the number of people participating in the network, and especially willing to share information with the network. We propose Graphsheets, based on Factsheets by \cite{arnold2019factsheets}, to provide standardized information, to increase trust in the contact tracing applications and the underlying GNN model predictions.

Finally, we draw on the \textit{nudge} idea introduced by \cite{thaler2009nudge} in behavioural sciences, to encourage users to share relevant information based on explainability techniques and graphsheets.

\section{Related Work}


\cite{tang2020privacy} presents a survey of currently available contact tracing applications and the technology choices. \cite{mit2020covidtracingtracker} maintains a Contract Tracing tracker apps with data on their adoption in respective geographical areas. \cite{mosoff2020gpaw} also maintain a similar list of apps.

\cite{hamilton2017inductive} introduced GraphSAGE (SAmple and AggreGatE) an inductive framework that leverages node feature information (ex: text attributes, node degrees) to efficiently generate node embeddings for previously unseen data or entirely new (sub)graphs. In this inductive framework, we learn a function that generates embeddings by sampling and aggregating features from a node’s local neighborhood. \cite{you2019position} proposed a Position Aware Graph Neural Network that significantly improves performance on the Link Prediction task over the Graph Convolutional Networks.

Much of the recent work on explanations are based on post-hoc models that try to approximate the prediction of complex models using interpretable models. \cite{vannur2020data} present post-hoc explanations of the links predicted by a Graph Neural Network by treating it as a classification problem. They present explanations using LIME (\cite{ribeiro2016should}) and SHAP (\cite{lundberg2017unified}). \cite{arya2020ai} introduced the AIX360 toolkit which has a number of explainability solutions that can be used for post-hoc explanation of graph models, if they can posed as approximated as interpretable models. \cite{agarwal2020neural} introduced Neural Additive Models which learn a model for each feature to increase the interpretability.

\section{Link Prediction for Contact Tracing}
\label{experiments}

\begin{figure*}[htb]
    \includegraphics[width=\linewidth]{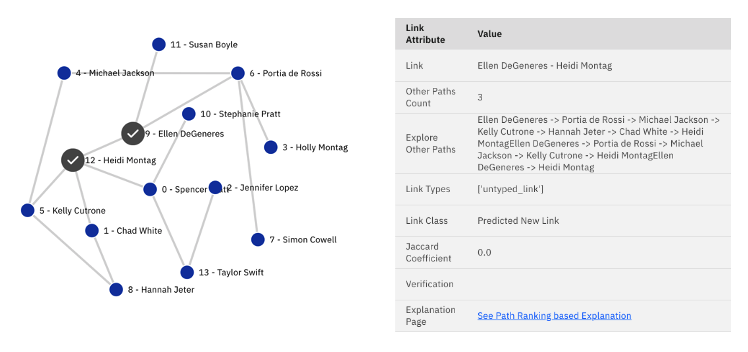}
    \caption{Example Graph which includes links predicted by a GNN model }
    \label{fig:example_graph}
\end{figure*}



We conducted our experiments on the UDBMS data \cite{UDMS_dataset} as a proxy for people in the real world. Each node in our graph has attributes similar to the personal data collected by the various Contact Tracing apps. A sample connected component is as shown in the Figure \ref{fig:example_graph}. In this particular example of the dataset, Ellen DeGeneres and Heidi Montag, do not seem to have any common neighbors or any shared node attributes, though they have paths involving other nodes. The challenge here is to explain links like these to the users of contact tracing applications.

We begin by explaining our experimental setup. We use the framework provided by \cite{you2019position}, which in turn uses pytorch \cite{paszke2019pytorch} and more specifically pytorch geometric \cite{fey2019fast}. One of the pre-processing steps we do is finding the all pairs shortest paths calculation using appropriate approximations. This pre-processing step comes in handy to explain the links as well.

Following the procedure in \cite{you2019position}, we choose only connected components with atleast 10 nodes for our experiments. A positive sample is created by randomly choosing 10\% of the links. For the negative sample, we use one of the nodes involved in the positive samples and pick a random unconnected node as the other node. The number of negative samples is same as that of the positive samples. We discuss hard negative samples in our future work section. Our batch size is typically 8 subgraphs and for PGNN, we use 64 anchor nodes.


\begin{table}[!htb]
    \begin{center}
    \begin{tabular}{cccc}
    \hline
    \textbf{Dataset} & \textbf{Model} & \textbf{ROC AUC} & \textbf{Std. Dev.} \\
    \hline
    {\textbf{UDBMS}}  &   \textbf{GCN} & 0.4689 & 0.0280   \\
                    & \textbf{P-GNN} & 0.6456 & 0.0185   \\
                        \cline{2-4}
    \hline
    \end{tabular}
    \caption{Comparison of Link Prediction performance with different GNNs}
    \label{tab:comparison}
    \end{center}
\end{table}

As shown on Table \ref{tab:comparison}, the P-GNN model seems to perform better on this dataset. In the next section, we'll explore ways to explain link prediction for the example shown in Figure \ref{fig:example_graph}.

\section{Explanations}

\begin{figure*}[htb]
    \includegraphics[width=\linewidth]{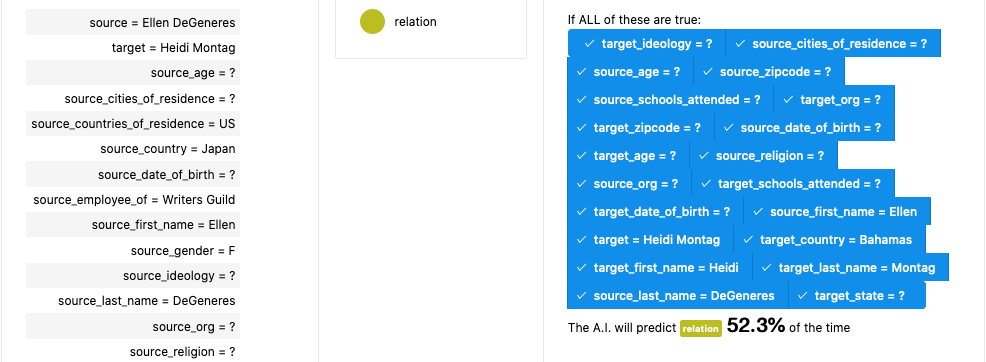}
    \caption{Explanations for Contact Tracing}
    \label{fig:anchors_explanations}
\end{figure*}

One of the commonly used methods for explainability is LIME \cite{ribeiro2016should}. Anchors \cite{ribeiro2018anchors} is an improvement over LIME which works by focusing only on the important features. The Link Prediction problem can perhaps be posed as a simple classification problem of predicting whether a person is exposed. A typical explanation for Link Prediction is as shown in Figure \ref{fig:anchors_explanations}. However, such interpretability solutions are rarely satisfactory to end users. Hence we propose an improvement to Anchors by incorporating ideas from GNN Explainer \cite{ying2019gnn}.

\subsection{Graph Anchors for Link Explanation}

We train a simple classifier to predict if a source and target nodes are linked or otherwise. This classifier is trained based on the output of the GNN model described in Section \ref{experiments}. Compared to the GNN model, this post hoc classification model is considered to be more interpretable.

In the above explanation, the classifier model is predicting a link between Ellen DeGeneres and Heidi Montag just like the original GNN model. Anchors model then explains that these two nodes will have a link 52.3\% of the time, if the conditions shown in Figure \ref{fig:anchors_explanations} hold. This result suffers from loss of information between the GNN model and the post-hoc interpretable model. Our classifier above is rudimentary and with more feature engineering, we could make the classifier model closer to the GNN model. However, even with reduced accuracy, providing a summary of the conditions under which a model predicts a link is more intuitive than the more complex graph explainability solutions.

\begin{figure*}[htb]
    \includegraphics[width=\linewidth]{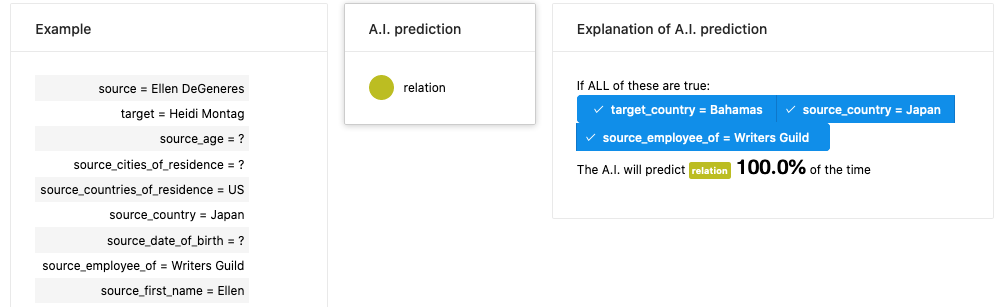}
    \caption{Anchors Explanations using GNN-Explainer output}
    \label{fig:graph_anchors_explanations}
\end{figure*}

Hence we tried tried generating Anchors explanation on the output of the GNN Explainer model, rather than the post-hoc classification model. The model consumed by the Anchors now consists of a subset of original features, and the target is a binary variable based on the output of the original GNN model. This simple modification to the input of the Anchors explainability solution seems to work reasonably but is still unsatisfactory from the end user point of view, as shown in Figure \ref{fig:graph_anchors_explanations}. GNN-Explainer and Anchors both use the broad idea or removing unnecessary features and using only the important features for explaining the links.

\subsection{Path Ranking based Link Explanation}

Our next explainability solution is inspired by ideas in Error detection in Knowledge Graphs. In particular, we use the PaTyBRED approach described in \cite{melo2017detection} and \cite{meloautomatic}.

\begin{figure*}[htb]
    \includegraphics[width=\linewidth]{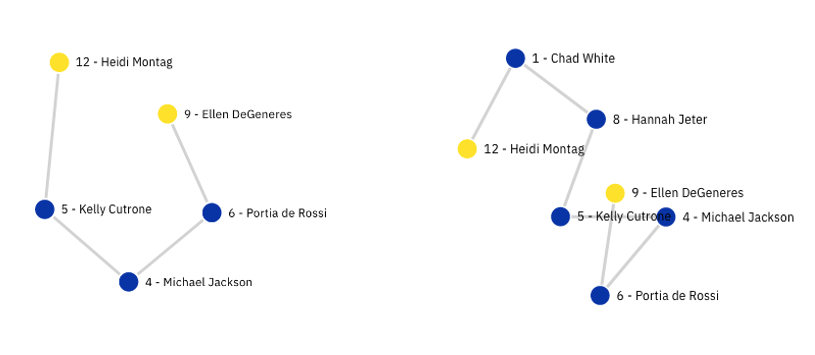}
    \caption{Path Ranking Algorithm based explanation}
    \label{fig:pbe_example}
\end{figure*}

We propose using the ranking function in the PaTyBRED algorithm to explain the link predicted by GNNs. The algorithm works by ranking already existing paths between two nodes. The idea here is that the information contained in that path is more useful than other paths to understand the predicted link. This idea is somewhat similar to the GNN Explainer idea of exploring the subgraph around the predicted link, except that we prefer to use an independent algorithm to rank all the paths rather than subgraph around the two nodes. As shown on Figure \ref{fig:pbe_example}, showing paths between two nodes that are predicted to be linked, can be a substitute for what would otherwise be unsatisfactory explanations using features and common neighbors.

\section{Graphsheets}

\begin{figure*}[htb]
    \includegraphics[width=\linewidth]{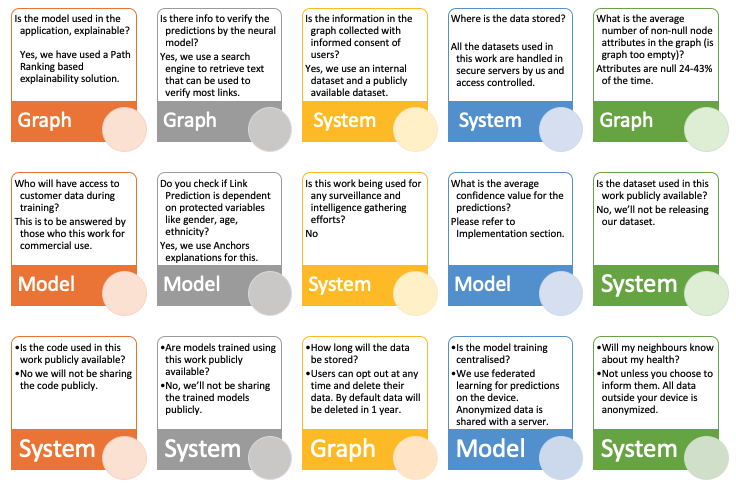}
    \caption{Graphsheet for Contact Tracing}
    \label{fig:graphsheets}
\end{figure*}

After having trained models for contact tracing use-case, it is important to ensure that downstream applications and model builders follow similar practices for fairness, privacy, data protection and ethical reasons. Towards this goal, we introduce Graph Sheets. Based on Factsheets \cite{arnold2019factsheets}, Graph Sheets include facts about the graph being studied and a number of FAQ types questions and answers that model developers have to consider before training Link Prediction models on their own datasets.

As shown in Figure \ref{fig:graphsheets}, we provide some sample information regarding the model, graph and the systems used in training Graph Neural Models on the dataset. Other relevant information can be added based on the application.

\section{Nudging}

Having created a Graph Sheet and being able to generate human understandable explanations for the predicted links, we can try to nudge users to both use contact tracing applications and to share their personal data with the community. \cite{thaler2009nudge} proposed \textit{nudging} to alter the choice architectures to make people behave in ways that are assumed to be good for them (without making other choices difficult or costly). In privacy preserving contact tracing, there is an inherent tension between the societal good for people to share personal information and the privacy risk involved with sharing such information. \cite{prainsack2020value} discusses the broader implications of nudging on public healthcare data. For the limited purpose of contact tracing where each participant is likely to benefit from sharing personal information, nudging them to share such information could be beneficial.

\begin{figure*}[htb]
\begin{subfigure}{0.48\textwidth}
    \includegraphics[width=\columnwidth]{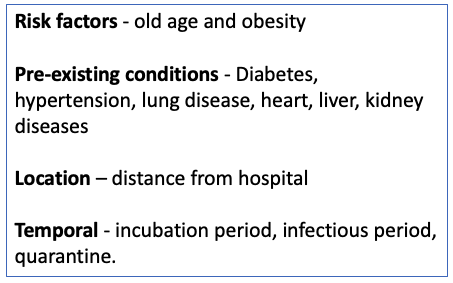}
    \caption{COVID19 Features}
    \label{fig:features}
\end{subfigure}
\begin{subfigure}{0.48\textwidth}
    \includegraphics[width=\columnwidth]{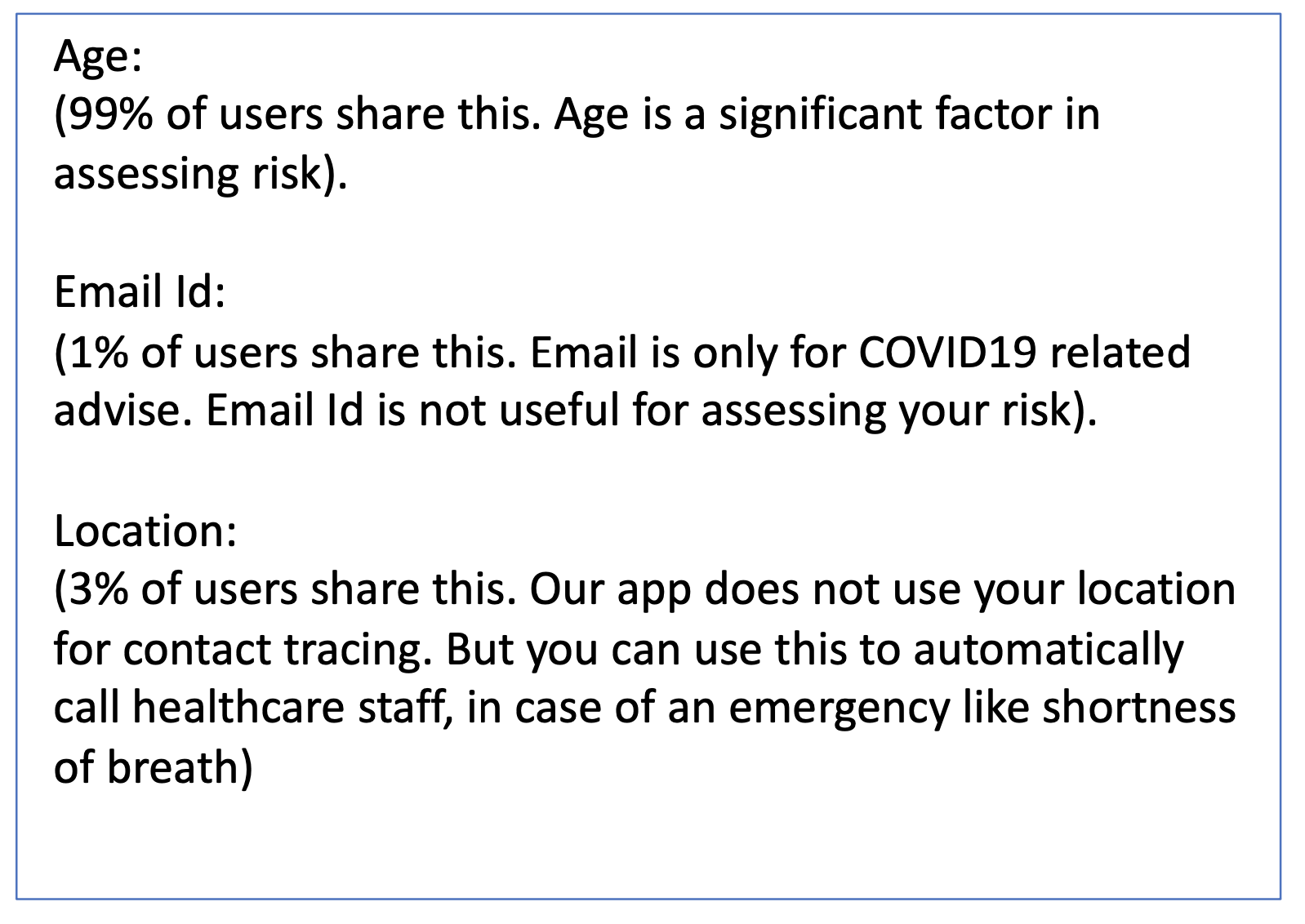}
    \caption{Nudging based on explanations}
    \label{fig:nudging}
\end{subfigure}
\caption{Users of contact tracing apps can be encouraged to share more information by explaining why the information is relevant}
\end{figure*}

As shown in Figure \ref{fig:nudging}, insights drawn from both explainability and graph sheets can be used for nudging. Email Id, is not a useful feature for contact tracing and hence is better not collected, or made clear to users that it's optional and used only for administrative purposes, as the case may be.

Age on the other hand, is a significant factor is assessing risk for COVID19 or other health factors. Models could be trained with age as a feature or alternatively, the age could be used to tailor alerts while being stored only on the user's device. For example, if the model predicts an exposure link with less confidence, the app could make a decision to alert the user or otherwise based on the risk profile of the user. 

\section*{Conclusion}

We described a Graph Neural Network model to predict exposure links in a contact tracing application to reduce false positives (unnecessary alerts) and also false negatives (lack of updates by participants). We proposed three methods to encourage people to both install and share information with the community, and optionally a centralized authority, by increasing trust in the applications. The three methods are namely, graph sheets to increase trust, human understandable explanations for the exposure notifications, and nudging based on important features found by explainability techniques.

\section*{Acknowledgements}

We thank the reviewers for their comments and feedback that helped improve this work.

\bibliographystyle{plainnat}
\bibliography{references}

\begin{thebibliography}{24}
\providecommand{\natexlab}[1]{#1}
\providecommand{\url}[1]{\texttt{#1}}
\expandafter\ifx\csname urlstyle\endcsname\relax
  \providecommand{\doi}[1]{doi: #1}\else
  \providecommand{\doi}{doi: \begingroup \urlstyle{rm}\Url}\fi

\bibitem[Abueg et~al.(2020)Abueg, Hinch, Wu, Liu, Probert, Wu, Eastham, Shafi,
  Rosencrantz, Dikovsky, et~al.]{abueg2020modeling}
Matthew Abueg, Robert Hinch, Neo Wu, Luyang Liu, William~JM Probert, Austin Wu,
  Paul Eastham, Yusef Shafi, Matt Rosencrantz, Michael Dikovsky, et~al.
\newblock Modeling the combined effect of digital exposure notification and
  non-pharmaceutical interventions on the covid-19 epidemic in washington
  state.
\newblock \emph{medRxiv}, 2020.

\bibitem[Agarwal et~al.(2020)Agarwal, Frosst, Zhang, Caruana, and
  Hinton]{agarwal2020neural}
Rishabh Agarwal, Nicholas Frosst, Xuezhou Zhang, Rich Caruana, and Geoffrey~E
  Hinton.
\newblock Neural additive models: Interpretable machine learning with neural
  nets.
\newblock \emph{arXiv preprint arXiv:2004.13912}, 2020.

\bibitem[Arnold et~al.(2019)Arnold, Bellamy, Hind, Houde, Mehta,
  Mojsilovi{\'c}, Nair, Ramamurthy, Olteanu, Piorkowski,
  et~al.]{arnold2019factsheets}
Matthew Arnold, Rachel~KE Bellamy, Michael Hind, Stephanie Houde, Sameep Mehta,
  A~Mojsilovi{\'c}, Ravi Nair, K~Natesan Ramamurthy, Alexandra Olteanu, David
  Piorkowski, et~al.
\newblock Factsheets: Increasing trust in ai services through supplier's
  declarations of conformity.
\newblock \emph{IBM Journal of Research and Development}, 63\penalty0
  (4/5):\penalty0 6--1, 2019.

\bibitem[Arya et~al.(2020)Arya, Bellamy, Chen, Dhurandhar, Hind, Hoffman,
  Houde, Liao, Luss, Mojsilovi{\'c}, et~al.]{arya2020ai}
Vijay Arya, Rachel~KE Bellamy, Pin-Yu Chen, Amit Dhurandhar, Michael Hind,
  Samuel~C Hoffman, Stephanie Houde, Q~Vera Liao, Ronny Luss, Aleksandra
  Mojsilovi{\'c}, et~al.
\newblock Ai explainability 360: hands-on tutorial.
\newblock In \emph{Proceedings of the 2020 Conference on Fairness,
  Accountability, and Transparency}, pages 696--696, 2020.

\bibitem[Bengio(2020)]{bengio2020b}
Yoshua Bengio.
\newblock A covid-19 exposure notification app won't work if people aren't
  motivated to use it: Ai expert.
\newblock
  \url{https://montrealgazette.com/health/a-contact-tracing-app-wont-work-if-people-arent-motivated-to-use-it-ai-expert},
  2020.
\newblock [Online; accessed 11-Oct-2020].

\bibitem[Bengio et~al.(2020)Bengio, Janda, Yu, Ippolito, Jarvie, Pilat, Struck,
  Krastev, and Sharma]{bengio2020need}
Yoshua Bengio, Richard Janda, Yun~William Yu, Daphne Ippolito, Max Jarvie, Dan
  Pilat, Brooke Struck, Sekoul Krastev, and Abhinav Sharma.
\newblock The need for privacy with public digital contact tracing during the
  covid-19 pandemic.
\newblock \emph{The Lancet Digital Health}, 2020.

\bibitem[Fey and Lenssen(2019)]{fey2019fast}
Matthias Fey and Jan~Eric Lenssen.
\newblock Fast graph representation learning with pytorch geometric.
\newblock \emph{arXiv preprint arXiv:1903.02428}, 2019.

\bibitem[Hamilton et~al.(2017)Hamilton, Ying, and
  Leskovec]{hamilton2017inductive}
Will Hamilton, Zhitao Ying, and Jure Leskovec.
\newblock Inductive representation learning on large graphs.
\newblock In \emph{Advances in neural information processing systems}, pages
  1024--1034, 2017.

\bibitem[Lu et~al.(2016)Lu, Chen, and Zhang]{UDMS_dataset}
Jiaheng Lu, Jun Chen, and Chao Zhang.
\newblock Helsinki {M}ulti-{M}odel {D}ata {R}epository.
\newblock http://udbms.cs.helsinki.fi/?dataset, 2016.

\bibitem[Lundberg and Lee(2017)]{lundberg2017unified}
Scott~M Lundberg and Su-In Lee.
\newblock A unified approach to interpreting model predictions.
\newblock In \emph{Advances in neural information processing systems}, pages
  4765--4774, 2017.

\bibitem[Melo and Paulheim(2017)]{melo2017detection}
Andr{\'e} Melo and Heiko Paulheim.
\newblock Detection of relation assertion errors in knowledge graphs.
\newblock In \emph{Proceedings of the Knowledge Capture Conference}, pages
  1--8, 2017.

\bibitem[Melo and Paulheim(2020)]{meloautomatic}
Andre Melo and Heiko Paulheim.
\newblock Automatic detection of relation assertion errors and induction of
  relation constraints.
\newblock \emph{Semantic Web}, 11\penalty0 (5):\penalty0 801--830, 2020.

\bibitem[MIT COVID Tracing Tracker {[2020]}()]{mit2020covidtracingtracker2}
MIT COVID Tracing Tracker {[2020]}.
\newblock {MIT Technology Review COVID Tracing Tracker Data}.
\newblock
  \url{https://docs.google.com/spreadsheets/d/1ATalASO8KtZMx__zJREoOvFh0nmB-sAqJ1-CjVRSCOw/edit#gid=0},
  2020.
\newblock [Online, accessed 12-Oct-2020].

\bibitem[MIT Technology Review {[2020]}()]{mit2020covidtracingtracker}
MIT Technology Review {[2020]}.
\newblock {MIT Technology Review COVID Tracing Tracker}.
\newblock
  \url{https://www.technologyreview.com/2020/05/07/1000961/launching-mittr-covid-tracing-tracker/},
  2020.
\newblock [Online, accessed 14-May-2020].

\bibitem[Mosoff et~al.(2020)Mosoff, Friedlich, Scassa, Bronson, and
  Millar]{mosoff2020gpaw}
Ryan Mosoff, Tommy Friedlich, Teresa Scassa, Kelly Bronson, and Jason Millar.
\newblock Global pandemic app watch (gpaw): Covid-19 exposure notification and
  contact tracing apps.
\newblock \url{https://craiedl.ca/gpaw/}, 2020.
\newblock [Online; accessed 11 October 2020].

\bibitem[Paszke et~al.(2019)Paszke, Gross, Massa, Lerer, Bradbury, Chanan,
  Killeen, Lin, Gimelshein, Antiga, et~al.]{paszke2019pytorch}
Adam Paszke, Sam Gross, Francisco Massa, Adam Lerer, James Bradbury, Gregory
  Chanan, Trevor Killeen, Zeming Lin, Natalia Gimelshein, Luca Antiga, et~al.
\newblock Pytorch: An imperative style, high-performance deep learning library.
\newblock In \emph{Advances in neural information processing systems}, pages
  8026--8037, 2019.

\bibitem[Prainsack(2020)]{prainsack2020value}
Barbara Prainsack.
\newblock The value of healthcare data: to nudge, or not?
\newblock \emph{Policy Studies}, pages 1--16, 2020.

\bibitem[Ribeiro et~al.(2016)Ribeiro, Singh, and Guestrin]{ribeiro2016should}
Marco~Tulio Ribeiro, Sameer Singh, and Carlos Guestrin.
\newblock " why should i trust you?" explaining the predictions of any
  classifier.
\newblock In \emph{Proceedings of the 22nd ACM SIGKDD international conference
  on knowledge discovery and data mining}, pages 1135--1144, 2016.

\bibitem[Ribeiro et~al.(2018)Ribeiro, Singh, and Guestrin]{ribeiro2018anchors}
Marco~Tulio Ribeiro, Sameer Singh, and Carlos Guestrin.
\newblock Anchors: High-precision model-agnostic explanations.
\newblock In \emph{AAAI}, volume~18, pages 1527--1535, 2018.

\bibitem[Tang(2020)]{tang2020privacy}
Qiang Tang.
\newblock Privacy-preserving contact tracing: current solutions and open
  questions.
\newblock \emph{arXiv preprint arXiv:2004.06818}, 2020.

\bibitem[Thaler and Sunstein(2009)]{thaler2009nudge}
Richard~H Thaler and Cass~R Sunstein.
\newblock \emph{Nudge: Improving decisions about health, wealth, and
  happiness}.
\newblock Penguin, 2009.

\bibitem[Vannur et~al.(2020)Vannur, Ganesan, Nagalapatti, Patel, and
  Thippeswamy]{vannur2020data}
Lingraj~S Vannur, Balaji Ganesan, Lokesh Nagalapatti, Hima Patel, and
  MN~Thippeswamy.
\newblock Data augmentation for personal knowledge base population, 2020.

\bibitem[Ying et~al.(2019)Ying, Bourgeois, You, Zitnik, and
  Leskovec]{ying2019gnn}
Rex Ying, Dylan Bourgeois, Jiaxuan You, Marinka Zitnik, and Jure Leskovec.
\newblock Gnn explainer: A tool for post-hoc explanation of graph neural
  networks.
\newblock \emph{arXiv preprint arXiv:1903.03894}, 2019.

\bibitem[You et~al.(2019)You, Ying, and Leskovec]{you2019position}
Jiaxuan You, Rex Ying, and Jure Leskovec.
\newblock Position-aware graph neural networks.
\newblock \emph{arXiv preprint arXiv:1906.04817}, 2019.

\end{thebibliography}

\end{document}